\documentclass[12pt]{spieman}
\usepackage{amsmath,amssymb,amsfonts}
\usepackage{graphicx}
\usepackage{overpic}
\usepackage{colortbl}
\usepackage{booktabs}
\usepackage{xcolor}
\usepackage{multirow}

\usepackage{soul}
\newcommand{\ms}[2]{\mbox{#1\,{\tiny$\pm$\,#2}}}
\makeatletter
\let\rev@SOUL@docmd\SOUL@docmd
\def\SOUL@docmd#1#2{%
  \ifx2#1%
    \SOUL@doword
    \def\SOUL@@##1##2{%
      \SOUL@token={#2{##1}{##2}}%
      \SOUL@everytoken
      \SOUL@syllable={#2{##1}{##2}}%
      \SOUL@everysyllable
      \SOUL@doword
      \SOUL@scan
    }%
  \else
    \def\SOUL@@{\rev@SOUL@docmd{#1}{#2}}%
  \fi
  \SOUL@@
}
\makeatother
\soulregister{\ms}{2}
\soulregister{\cite}{7}
\soulregister{\ref}{7}
\soulregister{\eqref}{7}
\DeclareRobustCommand{\rev}[1]{#1}
\newcommand{\rms}[2]{\rev{\ms{#1}{#2}}}

\begin{document}
\doublespacing
\title{\rev{Incorporating synthetic data from diffusion models supports echocardiography downstream tasks}}
\author[b]{Milad Yazdani\textsuperscript{\textdagger}}
\author[c]{Pooria Ashrafian\textsuperscript{\textdagger}}
\author[c]{Moein Heidari}
\author[b]{Alireza Rafiei Sardouei}
\author[c,d]{Dena Shahriari}
\author[e,f]{Ilker Hacihaliloglu\textsuperscript{*}}
\affil[b]{Department of Electrical and Computer Engineering, University of British Columbia, Vancouver, British Columbia, Canada}
\affil[c]{School of Biomedical Engineering, University of British Columbia, Vancouver, British Columbia, Canada}
\affil[d]{Department of Orthopaedics, University of British Columbia, Vancouver, British Columbia, Canada}
\affil[e]{Department of Radiology, University of British Columbia, Vancouver, British Columbia, Canada}
\affil[f]{Department of Medicine, University of British Columbia, Vancouver, British Columbia, Canada}

\maketitle
\begin{abstract}
\textbf{Purpose:} \rev{Large labeled echocardiography datasets are costly to collect. We use diffusion models to generate synthetic training data for segmentation and cardiac phase classification.}

\textbf{Approach:} \rev{We study three generation approaches on CAMUS: an unconditional image space model, a latent model conditioned on plain English prompts or abstract tokens, and a ControlNet model conditioned on an abstract token and segmentation mask. We evaluate synthesis with Fr\'echet Inception Distance (FID) and augmentation using segmentation and classification of end diastole and end systole. We also evaluate segmentation within CardiacUDC and directly from CAMUS to CardiacUDC.}

\textbf{Results:} \rev{Conditioning on an abstract token and segmentation mask gives the lowest mean FID, 1.5 compared with 3.2 for the Semantic Diffusion Model. Abstract token augmentation gives the highest classification accuracy, improving ResNet-18 from 83.8\% to 87.2\% and VGG-16 from 74.5\% to 78.7\%. Synthetic augmentation gives the highest mean Dice of 87.6\% for the CAMUS 2D U-Net and 84.6\% for the CardiacUDC nnU-Net. When nnU-Net models are trained on CAMUS and evaluated directly on CardiacUDC, augmentation raises mean Dice from 74.6\% to 75.2\%.}

\textbf{Conclusions:} \rev{Incorporating synthetic images into training gives modest gains in CAMUS phase classification, segmentation within CAMUS and CardiacUDC, and direct segmentation from CAMUS to CardiacUDC.}
\end{abstract}


{\noindent\footnotesize\textbf{\textdagger}Milad Yazdani and Pooria Ashrafian contributed equally to this work.\par}
{\noindent\footnotesize\textbf{*}Corresponding author: Ilker Hacihaliloglu, \href{mailto:ilker.hacihaliloglu@ubc.ca}{ilker.hacihaliloglu@ubc.ca}.\par}

\section{Introduction}
\label{sec:introduction}
Cardiovascular diseases pose a growing threat to global health and are a leading cause of death worldwide~\cite{roth2020global}. Early detection and accurate assessment of these conditions are critical, since many forms of cardiovascular disease can be managed or treated when identified in time. To support diagnosis, clinicians rely heavily on cardiac imaging techniques, which enable them to examine structural abnormalities, assess cardiac function, and monitor disease progression without resorting to invasive procedures. Among the various imaging modalities, echocardiography (echo) stands out as the most widely used for heart evaluation due to its accessibility, low cost, non-invasive acquisition, and ability to produce images in real time. Deep learning models~\cite{lecun2015deep} have become powerful tools for addressing persistent challenges in echocardiography, including the difficulty of interpreting noisy data and the low sensitivity and specificity often associated with clinical assessment. \rev{These models require large and diverse training datasets. The limited availability of high quality echo data restricts their development.} Recent work addresses this problem by using diffusion models to generate synthetic echo data for training deep learning models~\cite{olive2023synthetic}.

Since their inception, generative model architectures have garnered substantial interest in medical imaging~\cite{skandarani2023gans}. The pioneering generative adversarial networks (GANs)~\cite{goodfellow2014generative} have demonstrated strong versatility in generating high-quality medical samples efficiently. However, GANs are well known for issues such as poor coverage, mode collapse, and training instability. These limitations hinder accurate anatomical reproduction, especially when large structural variations are required from the conditional input, often resulting in noticeable anatomical collapse. Another generative approach is the Variational Autoencoder (VAE)~\cite{kingma2013auto}, which offers more stable training than GANs. However, VAEs typically struggle to generate high-quality samples due to their use of a simple Gaussian prior, which often results in blurry outputs. More recently, diffusion models~\cite{ho2020denoising} have emerged as state-of-the-art generative methods, showing greater robustness to the limitations of both GANs~\cite{dhariwal2021diffusion} and VAEs. Their strong performance makes them a promising alternative to GANs for medical image synthesis~\cite{mullerfranzes2023multimodal}.
Specifically, diffusion models have shown the capacity to produce high-quality samples with precise and detailed characteristics, which is crucial in medical image analysis where the generated samples must be both visually authentic and medically relevant~\cite{peng2023generating}. These models perform a forward diffusion phase where input data is incrementally perturbed with Gaussian noise, followed by learning to reverse this process to generate new samples from pure noise. Owing to their ability to learn complex data distributions, diffusion models excel in recognizing sharp and detailed features, particularly effective in medical images~\cite{khader2022medical}. However, the original denoising diffusion probabilistic model is unconditional~\cite{ho2020denoising}, so it cannot directly control the generated content. Targeting different generation cases can therefore require separate models, which increases training cost.

\rev{In this work, we study diffusion models conditioned on abstract tokens and segmentation masks to synthesize echo data.}

Our contributions are:
\begin{enumerate}
    \item \rev{We develop a diffusion framework that combines an abstract token with a segmentation map. The map controls annotated left heart geometry, while the token controls view dependent anatomy that is absent from the map, including the right atrium and right ventricle.}
    
    \item \rev{We create synthetic labeled image and mask pairs by sampling images from each training mask and retaining the conditioning mask as the label. This creates image and mask pairs without new manual annotations.}
\end{enumerate}

\section{Related Work}
\label{sec:related_work}
\subsection{Diffusion Models}
Diffusion models perform well across many domains and provide an alternative to generative adversarial networks~\cite{dhariwal2021diffusion,mullerfranzes2023multimodal,lin2023diffusion}. Recent work improves control over generated images and reduces computational cost. Stable Diffusion (SD)~\cite{rombach2022high} applies diffusion in a lower dimensional latent space to reduce memory use and computation. ControlNet~\cite{zhang2023adding} adds a parallel branch to the U-Net~\cite{ronneberger2015u} in SD. This branch incorporates conditions such as edge maps, segmentation maps, and key points. As an alternative, optimal transport flow matching reduces inference time for medical image synthesis while preserving image quality~\cite{yazdani2025flow}.

\subsection{Diffusion Models in Medical Applications}
Medical imaging applications of diffusion models include anomaly detection~\cite{iqbal2023unsupervised}, synthetic image generation~\cite{peng2023generating}, and segmentation~\cite{wolleb2022diffusion}. These models have also been applied to ultrasound dehazing and segmentation~\cite{stevens2024dehazing,yao2024dfcg}.

\subsection{Echo Synthesis via Diffusion Models}
Several studies use diffusion models to generate echocardiography images or videos~\cite{stojanovski2023echo,olive2023synthetic}. Stojanovski et al. use a Semantic Diffusion Model (SDM)~\cite{wang2022semantic} conditioned on semantic label maps. Synthetic Boost~\cite{adhikari2023synthetic} uses datasets from diffusion models to improve models that combine vision and language for echocardiography segmentation. Reynaud et al.~\cite{reynaud2023feature} propose a video diffusion model that generates sequences from single frames and clinical parameters.

\section{Methods}
\label{sec:methods}
\begin{figure}[!htbp]
    \centering
    \begin{overpic}[percent,width=\textwidth]{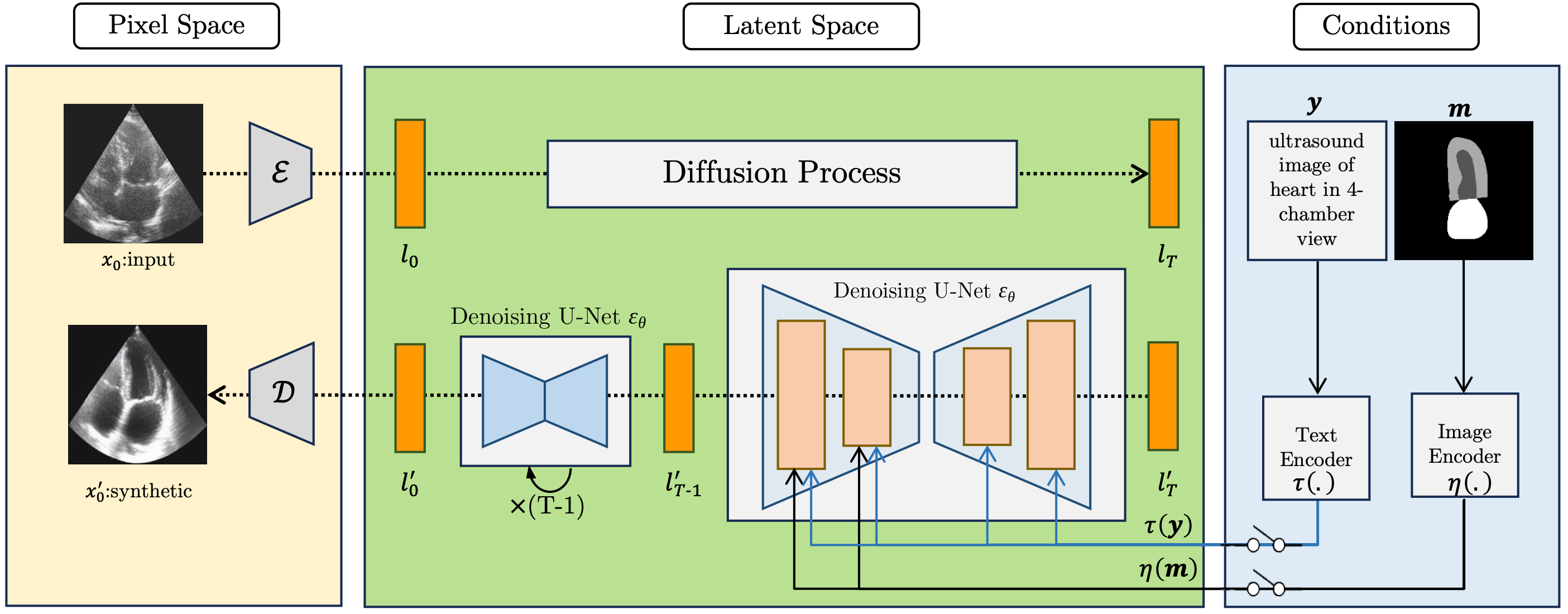}
    \end{overpic}
    \caption{
    \rev{Overview of the conditioned latent diffusion models. Both plain English prompts and abstract tokens use the fixed CLIP text encoder~\cite{radford2021learning}. For generation with an abstract token and segmentation mask, the image encoder block processes the map within the trainable ControlNet branch. The Stable Diffusion U-Net remains fixed.}}
    \label{fig:proposed_method}
\end{figure}

\rev{We study three generation approaches. The unconditional model in Section~\ref{subsec::unconditional_image_generation} is a denoising diffusion probabilistic model that operates in image space. The text conditioned model in Section~\ref{subsec::text_guided_image_generation} and the model conditioned on an abstract token and segmentation mask in Section~\ref{subsec::text_plut_segmentation_guided_image_generation} operate in the latent space of Stable Diffusion~\cite{rombach2022high}. For these two models, an encoder $\mathcal{E}$ maps an image $\mathbf{x}_0\in\mathbb{R}^{H\times W\times C}$ to $\mathbf{l}_0=\mathcal{E}(\mathbf{x}_0)$, and a decoder maps the denoised latent representation to an image. Figure~\ref{fig:proposed_method} illustrates the text conditioned model and the model conditioned on an abstract token and segmentation mask.}

\subsection{Unconditional Image Generation}\label{subsec::unconditional_image_generation}
\rev{For the unconditional setting, we use a standard denoising diffusion probabilistic model (DDPM)~\cite{ho2020denoising}. Let $\mathbf{x}_0$ denote a training image. At a uniformly sampled diffusion step $i \in \{1,\ldots,N\}$, where $N$ is the number of diffusion timesteps, the forward process produces}
\begin{equation}
    \mathbf{x}_i = \sqrt{\bar{\alpha}_i}\mathbf{x}_0 + \sqrt{1-\bar{\alpha}_i}\boldsymbol{\epsilon},
    \qquad \boldsymbol{\epsilon}\sim\mathcal{N}(\mathbf{0},\mathbf{I}),
\end{equation}
\rev{where $\mathbf{I}$ is the identity matrix, $\alpha_i=1-\beta_i$, $\bar{\alpha}_i=\prod_{j=1}^{i}\alpha_j$, and $\beta_i$ is the noise variance. A U-Net $\boldsymbol{\epsilon}_\theta$ is trained to predict the added noise using}
\begin{equation}
    \mathcal{L}_{\mathrm{DDPM}}
    =
    \mathbb{E}_{\substack{\mathbf{x}_0\sim p_{\mathrm{data}},\,
    i\sim\mathcal{U}(1,N),\,
    \boldsymbol{\epsilon}\sim\mathcal{N}(\mathbf{0},\mathbf{I})}}
    \left[\left\|\boldsymbol{\epsilon}-\boldsymbol{\epsilon}_\theta(\mathbf{x}_i,i)\right\|_2^2\right].
\end{equation}
\rev{During sampling, the model starts from Gaussian noise and applies the DDPM reverse steps to generate an image. Unlike the two conditioned settings, this model operates directly in image space. We train a separate model for each combination of view and cardiac phase.}

\subsection{Text Conditioned Image Generation}\label{subsec::text_guided_image_generation}
\rev{The text conditioned model uses Stable Diffusion~\cite{rombach2022high} as its backbone. The fixed CLIP text encoder $\tau(\cdot)$ maps a prompt $\mathbf{y}$ to an embedding that conditions the denoising U-Net through cross attention~\cite{radford2021learning}. At diffusion timestep $i$, the latent representation is}
\begin{equation}
    \mathbf{l}_i = \sqrt{\bar{\alpha}_i}\mathbf{l}_0
    + \sqrt{1-\bar{\alpha}_i}\boldsymbol{\epsilon},
    \qquad \boldsymbol{\epsilon}\sim\mathcal{N}(\mathbf{0},\mathbf{I}).
\end{equation}
\rev{We train the denoising U-Net $\boldsymbol{\epsilon}_\theta$ to predict the noise using}
\begin{equation}
    \mathcal{L}_{\mathrm{text}}
    =
    \mathbb{E}
    \left[
    \left\|
    \boldsymbol{\epsilon}
    -\boldsymbol{\epsilon}_\theta(\mathbf{l}_i,i,\tau(\mathbf{y}))
    \right\|_2^2
    \right].
\end{equation}
\rev{The expectation is over training image and prompt pairs, diffusion timesteps, and noise. The VAE and CLIP text encoder remain fixed. We train separate instances with plain English prompts and abstract tokens. Section~\ref{subsec:Prompt_Engineering} defines both representations.}

\subsection{Generation with an Abstract Token and Segmentation Mask}\label{subsec::text_plut_segmentation_guided_image_generation}

\rev{This model adds a segmentation map $\mathbf{m}$ through ControlNet~\cite{zhang2023adding}. A trainable map encoder $\eta(\cdot)$ processes the map. ControlNet copies the encoder and middle blocks of the Stable Diffusion U-Net into a trainable branch and connects this branch to the fixed U-Net through zero initialized convolutions. The noise prediction objective becomes}
\begin{equation}
    \mathcal{L}_{\mathrm{token+seg}}
    =
    \mathbb{E}
    \left[
    \left\|
    \boldsymbol{\epsilon}
    -\boldsymbol{\epsilon}_{\theta,\phi}
    (\mathbf{l}_i,i,\tau(\mathbf{y}),\eta_\phi(\mathbf{m}))
    \right\|_2^2
    \right],
\end{equation}
\rev{where $\phi$ denotes the trainable ControlNet parameters and $\theta$ denotes the fixed Stable Diffusion U-Net parameters. The expectation is over training images, abstract tokens, corresponding maps, diffusion timesteps, and noise. The VAE and CLIP text encoder also remain fixed.}

\subsection{Prompt Engineering}\label{subsec:Prompt_Engineering}
\rev{The text conditioned model encodes the imaging modality, organ, view, and cardiac phase in each prompt. The plain English prompt uses the corresponding words. The abstract token prompt replaces each word with a fixed string of six uppercase letters. We generate the mapping with seed 0 and keep it fixed during training and sampling. WZTEJD denotes ultrasound, XCVKPR denotes the heart, MYNBIQ denotes 2CH, PMZJPL denotes 4CH, SGQEJE denotes ED, and YDTZIR denotes ES.}

\rev{For generation with an abstract token and segmentation mask, we use one fixed abstract token for 2CH and another for 4CH. The token encodes the view but not the cardiac phase. The segmentation map supplies the annotated geometry associated with the phase.}

\section{Experiments}
\label{sec:Experiments}

\subsection{Experimental Data}
\label{subsec:Experimental_Data}
We use the CAMUS echocardiography dataset~\cite{leclerc2019deep}. It contains two dimensional apical two chamber (2CH) and four chamber (4CH) views from 500 patients at end diastole (ED) and end systole (ES). CAMUS is public and contains deidentified data. \rev{The official CAMUS test split contains 50 patients without segmentation masks.} Following Stojanovski et al.~\cite{stojanovski2023echo}, we use 400 patients for training and 50 patients for validation. This gives 1600 training images and 200 validation images. The split is at the patient level. We evaluate CAMUS generation, segmentation, and classification on the validation split.

\rev{We also use 2257 annotated frames from the public apical four chamber (A4C) release of CardiacUDA, named CardiacUDC in the release~\cite{yang2023graphecho}. The masks label the left ventricle (LV), left atrium (LA), right atrium (RA), and right ventricle (RV). We split the data by recording group. Near duplicate recordings remain in the same group and split. The training split has 1804 frames from 229 groups. The validation split has 226 frames from 29 groups. The test split has 227 frames from 28 groups. For the within dataset experiments, the generator and segmentation models are trained on the CardiacUDC training split. The validation and test splits consist of real frames.}

\subsection{Implementation and Statistical Analysis}
\label{subsec:Experimental_Settings}
For the text conditioned models, we train the denoising U-Net of Stable Diffusion v1.4 at $512 \times 512$ resolution. The batch size is 1, with gradient accumulation over 4 batches. We use gradient checkpointing, exponential moving average, and AdamW with a constant learning rate of $1\mathrm{e}{-5}$. The VAE and CLIP text encoder remain fixed. \rev{The text conditioned models with plain English prompts and abstract tokens use the same Stable Diffusion v1.4 denoising U-Net and training protocol described above.} \rev{For generation with an abstract token and segmentation mask, we train only ControlNet on top of Stable Diffusion v1.5 at $512 \times 512$ resolution.} ControlNet training uses four NVIDIA V100 GPUs, each with 16 GB of memory. We use 120000 optimizer updates, batch size 1, gradient checkpointing, 8 bit Adam, and a learning rate of $5\mathrm{e}{-6}$. The Stable Diffusion components remain fixed. All diffusion models use a noise schedule with 1000 diffusion timesteps. At sampling, the text conditioned models use a UniPC scheduler~\cite{zhao2024unipc} with 40 denoising steps. \rev{The model with an abstract token and segmentation mask uses 50 denoising steps.}

For unconditional generation, we train one image space DDPM for each view and phase combination. Each model is trained for 400 epochs on grayscale images resized to $128 \times 128$ and normalized to $[-1,1]$. Training uses random horizontal flips. The U-Net has six downsampling and six upsampling blocks with channel sizes $(64,64,128,128,256,256)$. It has two ResNet layers per block and one self attention block in each path. We use AdamW with batch size 16, an initial learning rate of $1\mathrm{e}{-4}$, a cosine schedule with 500 warmup steps, and seed 0. Sampling uses all 1000 DDPM reverse steps.

For image synthesis, we compute Fr\'echet Inception Distance (FID)~\cite{heusel2017gans} and Kernel Inception Distance (KID)~\cite{binkowski2018demystifying} between generated and real images. FID compares the distributions of image features. KID measures their squared maximum mean discrepancy. \rev{We set the Inception v3 feature dimension to 64 for both metrics and report KID without additional scaling. For each system, view, phase, and sampling seed, we compare 50 generated images with the same 50 real validation images. We report the mean and sample standard deviation (SD) over five sampling seeds from one fixed checkpoint for each model.}

For CAMUS segmentation, we use a 2D U-Net~\cite{ronneberger2015u} with four output classes and channel sizes $(2,4,8,16,32,64,128,256)$. Images are resized to $256 \times 256$. Training uses Adam, cosine annealing, and the checkpoint with the highest mean validation Dice. \rev{Every training data condition uses 300 epochs and batch size 4. Checkpoint selection and reporting both use the CAMUS validation split. Each synthetic pair starts from a mask in the CAMUS training split. We use the mask to condition image generation and retain it as the segmentation label. Real+50\%, Real+100\%, and Real+200\% add 800, 1600, and 3200 synthetic pairs, respectively. Different noise seeds can generate multiple images from one mask. All generated image and mask pairs are retained without manual verification, screening, or filtering.}

For ED and ES classification, we use ImageNet pretrained ResNet-18 and VGG-16. Images are resized to $224 \times 224$. We replace the final layer with a two class head and leave about two million parameters trainable. Training runs for 30 epochs with SGD, momentum 0.9, a learning rate of 0.001, and batch size 32. We add up to 800 synthetic images per class to keep the training set balanced. Every synthetic condition denotes real training data augmented with images from the named generator.

For downstream segmentation, Dice measures overlap between discrete predicted and reference masks. Hausdorff distance (HD) is the maximum distance between their contours. Average surface distance (ASD) is the mean contour distance. HD and ASD are in pixels at $256 \times 256$ resolution. For classification, we report accuracy (ACC) and macro precision (PR), recall (RC), and F1. \rev{We average the results over five paired runs with different random initializations for each downstream comparison. For nnU-Net, we use five repetitions paired by run index. We report the sample SD for mean CAMUS Dice. We test paired differences with a two sided Student t test at $\alpha=0.05$ with four degrees of freedom. We report $p$ values.}

\rev{For CardiacUDC, we resize frames to $256 \times 256$ with bilinear interpolation and masks with nearest neighbor interpolation. The 2D U-Net has five output classes for the background and four cardiac structures. We train for 300 epochs with batch size 16, Adam with a learning rate of 0.001, and cosine annealing to $1\mathrm{e}{-4}$. The loss weight is 0.05 for the background and 0.25 for each cardiac structure. Training augmentation uses random affine, brightness, and contrast transformations. For each seed from 0 through 4, we select the checkpoint with the highest mean validation Dice and evaluate it on the test split. In the Real+50\%, Real+100\%, and Real+200\% conditions, synthetic frames equal 50\%, 100\%, and 200\% of the real training set, respectively. We train a separate generator with an abstract token and segmentation mask on the CardiacUDC training split. Its inputs are a segmentation map with four cardiac structures and a fixed abstract token for A4C.}

\rev{We also use nnU-Net~\cite{isensee2021nnu} version 2.8.1 in its 2D configuration. We generate the plan from the Real training frames and reuse it for Real+100\%. The plan uses a $224 \times 256$ patch and batch size 56. We use the default 2D network, loss, optimizer, schedule, and augmentation. Training runs for 1000 epochs with 250 updates per epoch. Real+100\% adds 1804 synthetic training frames. We repeat the predefined split five times, so run indices denote repetitions rather than cross validation folds. For each run, we select the checkpoint with the largest moving average of mean foreground Dice and evaluate it once on the 227 test frames.}

\subsection{Image Synthesis Evaluation}
\label{subsec:Image_Synthesis_Evaluation}
\rev{We compare five generation settings with the CAMUS validation images for each view and phase combination. The settings differ in architecture and conditioning input. Unconditional generation uses no input. The plain English and abstract token settings use text alone. SDM~\cite{stojanovski2023echo} uses a segmentation mask alone. Our joint setting uses an abstract token and segmentation mask. We report FID and KID for each view and phase and their mean. Section~\ref{subsec:Experimental_Settings} defines the common evaluation protocol and the training and sampling procedures for our models.}
\begin{table}[!htbp]
\centering
\caption{\rev{Comparison of generation systems and prompt representations using FID and KID. Values are means across generation seeds, and the value after $\pm$ is SD. Bold indicates the best results.}}
\label{tab:inception_scores}
{\footnotesize
\setlength{\tabcolsep}{1pt}
\begin{tabular}{@{}>{\raggedright\arraybackslash}p{0.30\textwidth}cccc|c|cccc|c@{}}
\toprule
\textbf{Method} & \multicolumn{5}{c}{\textbf{FID} $\downarrow$} & \multicolumn{5}{c}{\textbf{KID} $\downarrow$} \\
\cmidrule(lr){2-6} \cmidrule(lr){7-11}
& {\scriptsize\textbf{2CH-ED}} & {\scriptsize\textbf{2CH-ES}} & {\scriptsize\textbf{4CH-ED}} & {\scriptsize\textbf{4CH-ES}} & {\scriptsize\textbf{Mean}} & {\scriptsize\textbf{2CH-ED}} & {\scriptsize\textbf{2CH-ES}} & {\scriptsize\textbf{4CH-ED}} & {\scriptsize\textbf{4CH-ES}} & {\scriptsize\textbf{Mean}} \\
\midrule
Unconditional & \rms{4.8}{0.2} & \rms{4.6}{0.2} & \rms{5.2}{0.2} & \rms{4.1}{0.2} & \rms{4.7}{0.2} & \rms{1.5}{0.2} & \rms{2.6}{0.2} & \rms{2.6}{0.2} & \rms{2.0}{0.2} & \rms{2.2}{0.2} \\
\rev{Plain English} & \rms{1.6}{0.1} & \rms{2.5}{0.2} & \rms{1.9}{0.1} & \rms{3.0}{0.2} & \rms{2.3}{0.1} & \rms{1.4}{0.1} & \rms{2.6}{0.2} & \rms{1.7}{0.1} & \rms{2.7}{0.2} & \rms{2.1}{0.1} \\
\rev{Abstract token} & \rms{1.5}{0.1} & \rms{2.3}{0.1} & \rms{1.8}{0.1} & \rms{2.9}{0.2} & \rms{2.1}{0.1} & \textbf{\rms{1.3}{0.1}} & \rms{2.5}{0.2} & \textbf{\rms{1.6}{0.1}} & \rms{2.7}{0.2} & \rms{2.0}{0.2} \\
SDM & \rms{1.8}{0.1} & \rms{1.8}{0.1} & \rms{4.5}{0.2} & \rms{4.5}{0.2} & \rms{3.2}{0.2} & \rms{2.8}{0.2} & \rms{2.9}{0.2} & \rms{5.4}{0.3} & \rms{5.4}{0.3} & \rms{4.1}{0.3} \\
\rev{Abstract token and segmentation} & \textbf{\rms{1.4}{0.1}} & \textbf{\rms{1.6}{0.1}} & \textbf{\rms{1.6}{0.1}} & \textbf{\rms{1.3}{0.1}} & \textbf{\rms{1.5}{0.1}} & \textbf{\rms{1.3}{0.1}} & \textbf{\rms{1.5}{0.1}} & \textbf{\rms{1.6}{0.1}} & \textbf{\rms{1.6}{0.1}} & \textbf{\rms{1.5}{0.1}} \\
\bottomrule
\end{tabular}
}
\end{table}

\rev{Abstract token and segmentation has the lowest mean FID at 1.5, compared with 3.2 for SDM.}

\subsection{Qualitative Comparison Across Generation Systems}\label{subsec:Qualitative_Comparison_Across_Conditioning_Strategies}
\begin{figure}[!htbp]
    \centering
    \includegraphics[width=0.80\textwidth]{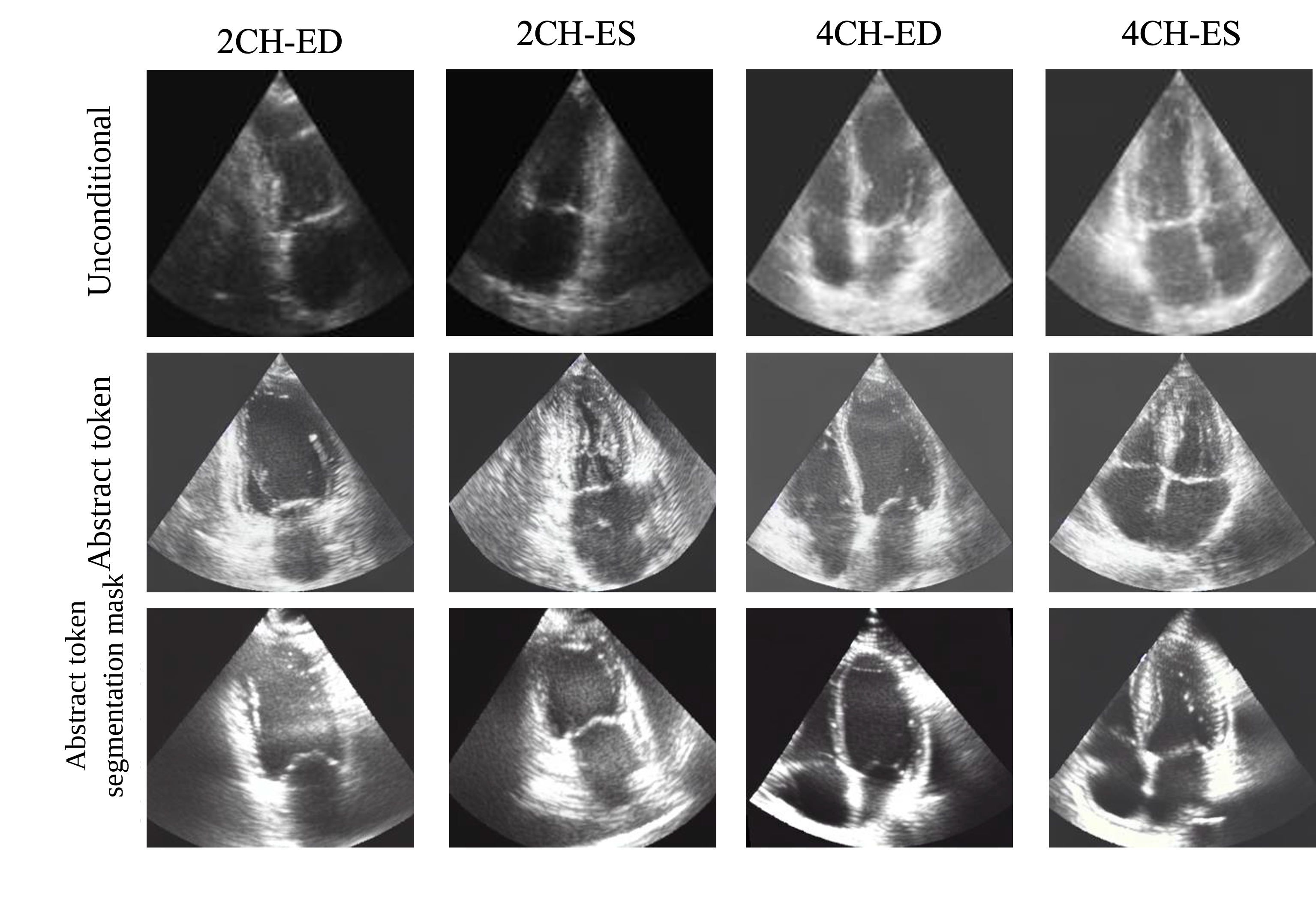}
    \caption{\rev{Synthetic examples from the three generation settings. Rows from top to bottom show unconditional generation, generation with an abstract token, and generation with an abstract token and segmentation mask. Columns from left to right show 2CH-ED, 2CH-ES, 4CH-ED, and 4CH-ES.}}
	\label{fig:conditioning_qualitative}
\end{figure}

Figure~\ref{fig:conditioning_qualitative} compares one synthetic image for each CAMUS view and phase combination under three generation settings. The rows show unconditional generation, generation with an abstract token, and generation with an abstract token and segmentation mask. The columns show 2CH-ED, 2CH-ES, 4CH-ED, and 4CH-ES.
In these examples, the unconditional outputs have lower global brightness in both 2CH columns and a partial mirror artifact in the 4CH-ES column. The outputs conditioned on an abstract token and segmentation mask show visible chamber boundaries in all four columns.

\begin{figure}[!htbp]
    \centering
    \includegraphics[width=0.56\textwidth]{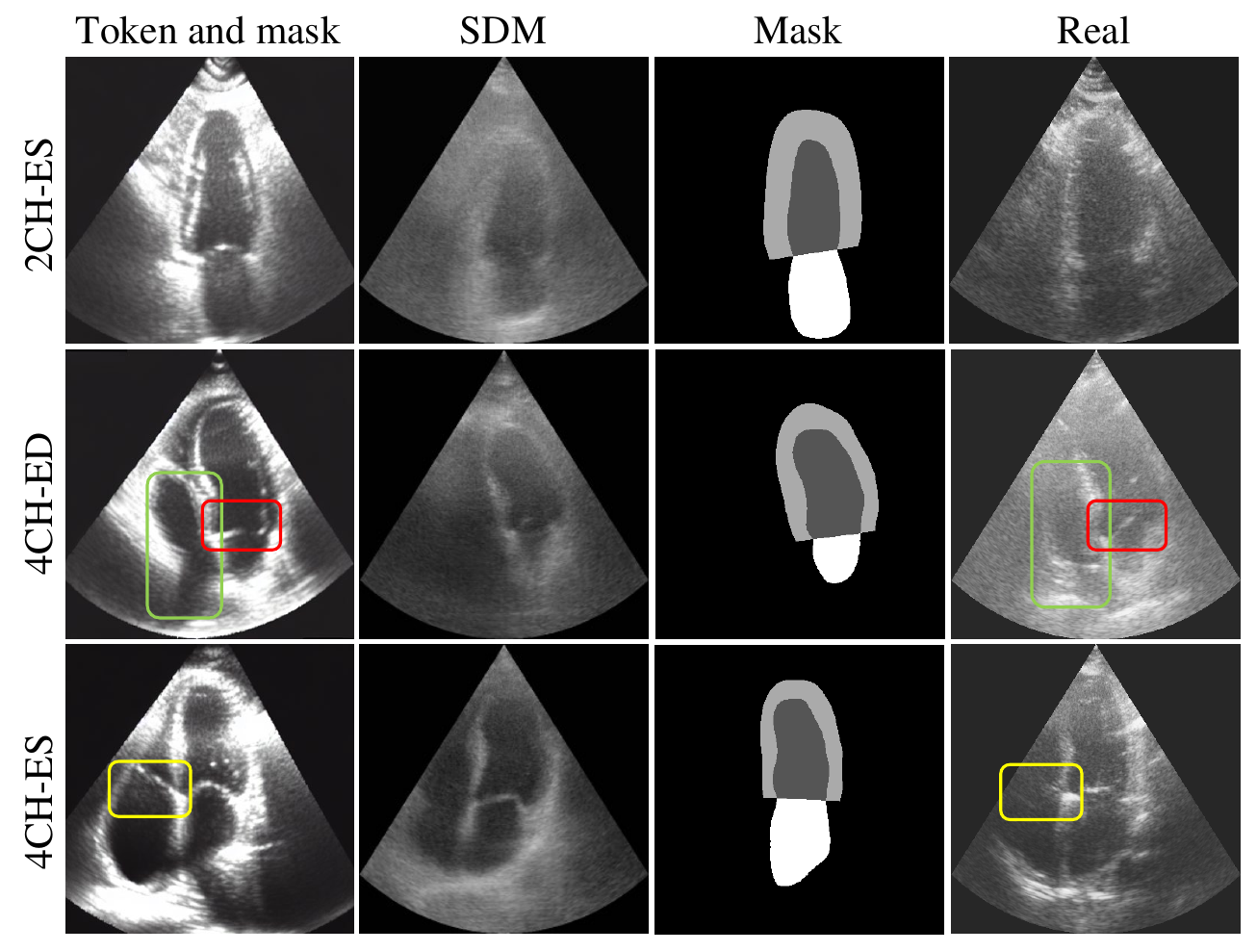}
    \caption{\rev{Generation results from the model conditioned on an abstract token and segmentation mask and from SDM~\cite{stojanovski2023echo}. Columns from left to right show the output from our model, SDM output, segmentation map, and real image. Rows show 2CH-ES, 4CH-ED, and 4CH-ES. Green, red, and yellow boxes mark the right chambers, mitral valve, and tricuspid valve.}}
		\label{fig:generation_result}
\end{figure}

Figure~\ref{fig:generation_result} compares the two map guided systems with the corresponding segmentation maps and real CAMUS images. The rows show 2CH-ES, 4CH-ED, and 4CH-ES examples.
In the two 4CH examples, the outputs conditioned on an abstract token and segmentation mask include visible right chambers, while the corresponding SDM outputs do not. The highlighted valve regions are also visible in these outputs and the real images.

\subsection{Prompt Representation Ablation}\label{subsec:Prompt_Ablation}
\rev{We compare Plain English and abstract token prompts using the same text conditioned architecture and training protocol. The synthesis comparison uses mean FID and mean KID across the four view and phase combinations. The downstream comparison uses ACC on the CAMUS validation split for ResNet-18 and VGG-16. Each prompt row uses real training data augmented with images from the corresponding generator. FID and KID require generated images, so these entries are not applicable for Real.}

\begin{table}[!htbp]
\centering
\caption{\rev{Prompt representation ablation using mean FID, mean KID, and mean ACC. Each $p$ column reports the paired two sided Student t test for ACC. Plain English is compared with Real, and Abstract token is compared with Plain English.}}
\label{tab:prompt-ablation}
{\footnotesize
\setlength{\tabcolsep}{3pt}
\begin{tabular}{@{}lcccccc@{}}
\toprule
\multirow{2}{*}{\textbf{Prompt}} & \multirow{2}{*}{\textbf{FID} $\downarrow$} & \multirow{2}{*}{\textbf{KID} $\downarrow$} & \multicolumn{2}{c}{\textbf{ResNet-18}} & \multicolumn{2}{c}{\textbf{VGG-16}} \\
\cmidrule(lr){4-5} \cmidrule(lr){6-7}
 & & & \textbf{ACC (\%)} $\uparrow$ & \textbf{$p$} & \textbf{ACC (\%)} $\uparrow$ & \textbf{$p$} \\
\midrule
\rev{Real} & \multicolumn{2}{c}{\rev{N/A}} & \rev{83.8} & \rev{N/A} & \rev{74.5} & \rev{N/A} \\
\rev{Plain English} & \rev{2.3} & \rev{2.1} & \rev{86.6} & \rev{$<0.01$} & \rev{78.0} & \rev{$<0.01$} \\
\rev{Abstract token} & \rev{2.1} & \rev{2.0} & \rev{87.2} & \rev{0.11} & \rev{78.7} & \rev{0.08} \\
\bottomrule
\end{tabular}
}
\end{table}

\rev{Abstract token prompting gives a mean FID of 2.1 and mean KID of 2.0, compared with 2.3 and 2.1 for plain English. Plain English increases ACC over Real by 2.8 points for ResNet-18 and 3.5 points for VGG-16. Abstract token ACC is 0.6 points higher for ResNet-18 and 0.7 points higher for VGG-16 than plain English ACC, with $p=0.11$ and $p=0.08$.}

\subsection{Contribution of Conditioning Inputs}\label{subsec:Text_Condition_Contribution}
\rev{We vary the sampling inputs of one model trained with abstract tokens and segmentation masks. The three conditions use the segmentation mask alone, the matching abstract token with the mask, or a nonmatching abstract token with the same mask. For each condition, we generate one image from each of 200 validation masks under each of five sampling seeds. This gives 1000 images, including 500 whose reference view is 4CH. Correct view rate is the percentage that matches the reference view of the source mask. Right chamber presence is the percentage of the 4CH images with both the RA and RV. Two readers who are blinded to the sampling condition assess these outcomes independently and resolve disagreements by consensus. Mask agreement Dice is the mean across all images. It compares each conditioning map with the three structures predicted by a fixed CAMUS 2D U-Net trained with Real from seed 0.}

\begin{table}[!htbp]
\centering
\caption{\rev{Contribution of the conditioning inputs using correct view rate, right chamber presence in 4CH, and mask agreement Dice. Values are aggregate percentages.}}
\label{tab:text-contribution}
{\footnotesize
\setlength{\tabcolsep}{5pt}
\begin{tabular}{@{}>{\raggedright\arraybackslash}p{0.36\textwidth}ccc@{}}
\toprule
\textbf{Condition} & \textbf{Correct view (\%)} $\uparrow$ & \textbf{Right chambers (\%)} $\uparrow$ & \textbf{Dice (\%)} $\uparrow$ \\
\midrule
\rev{Segmentation mask only} & \rev{82} & \rev{68} & \rev{86.3} \\
\rev{Matching abstract token and segmentation mask} & \rev{94} & \rev{90} & \rev{86.7} \\
\rev{Nonmatching abstract token and segmentation mask} & \rev{43} & \rev{39} & \rev{85.9} \\
\bottomrule
\end{tabular}
}
\end{table}

\rev{The matching abstract token increases correct view rate from 82\% to 94\% and right chamber presence from 68\% to 90\% compared with the segmentation mask alone. With a nonmatching abstract token, these rates are 43\% and 39\%, while mask agreement Dice ranges from 85.9\% to 86.7\% across the three conditions.}

\subsection{CAMUS Segmentation Assessment}\label{subsec:Downstream_tasks}
We compare a 2D U-Net trained with real CAMUS images against models trained with additional synthetic pairs. Real+50\%, Real+100\%, and Real+200\% use pairs from the generator conditioned on an abstract token and segmentation mask. The data sizes and training protocol are defined in Section~\ref{subsec:Experimental_Settings}. \rev{Checkpoint selection and reporting both use the 50 patient validation split defined by Stojanovski et al.~\cite{stojanovski2023echo}. The SDM comparison uses the Semantic Diffusion Model baseline~\cite{stojanovski2023echo}.}

\begin{table}[!htbp]
\centering
\caption{\rev{CAMUS segmentation results using Dice, HD, and ASD. Values are means across five seeds. Mean Dice is followed by its sample SD. The $p$ column compares mean Dice for each condition with Real using the paired two sided Student t test. HD and ASD are measured in pixels at $256 \times 256$ resolution.} Bold indicates the best results.}
\label{tab:dice-hd-asd}
{\footnotesize
\setlength{\tabcolsep}{2pt}
\resizebox{\textwidth}{!}{%
\begin{tabular}{@{}lccccc|cccc|cccc@{}}
\toprule
\multirow{2}{*}{\textbf{Training data}} & \multicolumn{5}{c|}{\textbf{Dice (\%)} $\uparrow$} & \multicolumn{4}{c|}{\textbf{HD} $\downarrow$} & \multicolumn{4}{c}{\textbf{ASD} $\downarrow$} \\
\cmidrule(lr){2-6} \cmidrule(lr){7-10} \cmidrule(lr){11-14}
 & \textbf{Mean} & \textbf{$p$} & \textbf{LV-endo} & \textbf{LV-epi} & \textbf{LA} & \textbf{Mean} & \textbf{LV-endo} & \textbf{LV-epi} & \textbf{LA} & \textbf{Mean} & \textbf{LV-endo} & \textbf{LV-epi} & \textbf{LA} \\
\midrule
Real & \rms{86.9}{0.3} & \rev{N/A} & \rev{89.0} & \rev{92.1} & \rev{79.6} & \rev{15.0} & \rev{9.6} & \rev{10.9} & \rev{24.5} & \rev{5.4} & \rev{3.8} & \rev{4.0} & \rev{8.3} \\
Real+50\% & \rms{87.2}{0.3} & \rev{0.01} & \rev{89.4} & \rev{92.3} & \rev{79.9} & \rev{12.1} & \textbf{\rev{9.2}} & \textbf{\rev{9.7}} & \rev{17.3} & \rev{4.9} & \textbf{\rev{3.7}} & \textbf{\rev{3.7}} & \rev{7.2} \\
Real+100\% & \textbf{\rms{87.6}{0.3}} & \rev{$<0.01$} & \textbf{\rev{89.9}} & \textbf{\rev{92.5}} & \rev{80.4} & \rev{12.5} & \rev{9.3} & \rev{10.4} & \rev{17.7} & \textbf{\rev{4.8}} & \rev{3.8} & \rev{3.8} & \rev{6.9} \\
Real+200\% & \rms{87.5}{0.5} & \rev{0.02} & \textbf{\rev{89.9}} & \rev{92.3} & \rev{80.2} & \textbf{\rev{11.8}} & \rev{9.4} & \rev{10.0} & \textbf{\rev{16.0}} & \rev{4.9} & \rev{3.9} & \rev{3.8} & \rev{6.9} \\
SDM & \rms{85.7}{0.4} & \rev{$<0.01$} & \rev{86.5} & \rev{90.0} & \textbf{\rev{80.6}} & \rev{13.7} & \rev{12.7} & \rev{11.6} & \rev{16.9} & \rev{5.2} & \rev{4.6} & \rev{4.6} & \textbf{\rev{6.5}} \\
\bottomrule
\end{tabular}%
}
}
\end{table}

\rev{Mean Dice reaches 87.6\% with Real+100\% and is 87.5\% with Real+200\%, compared with 86.9\% with Real. For Real+100\%, the LA accounts for 89.5\% of the net HD reduction across the three structures.}

\begin{figure}[!htbp]
    \centering
    \includegraphics[width=0.42\textwidth]{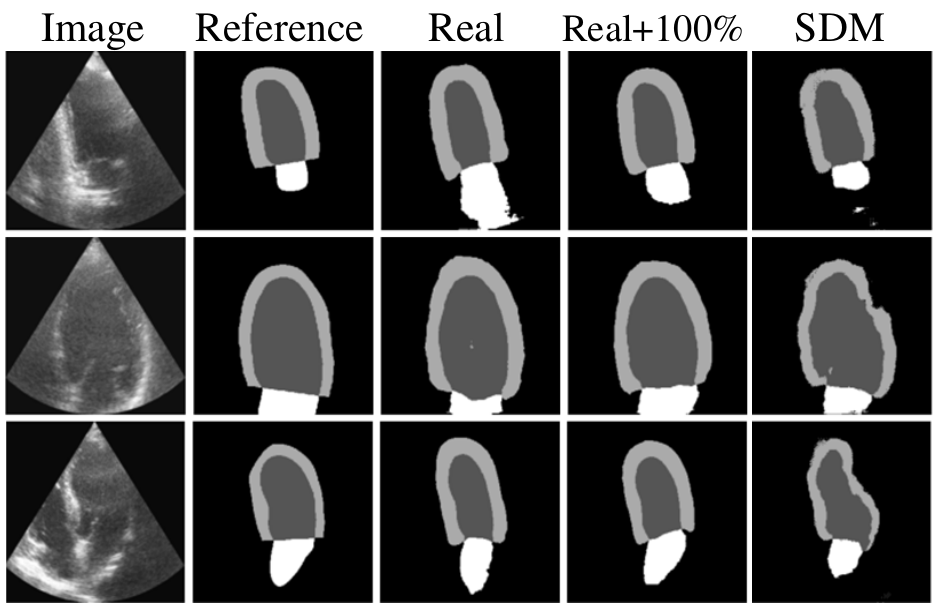}
    \caption{\rev{CAMUS segmentation examples. Columns from left to right show the image, reference mask, Real prediction, Real+100\% prediction, and SDM prediction.}}
		\label{fig:seg-vis}
\end{figure}

Figure~\ref{fig:seg-vis} compares three segmentation examples under Real, Real+100\%, and SDM training.
In these examples, the Real+100\% predictions follow the displayed reference contours. The SDM predictions show boundary deviations in the second and third rows.

\subsection{CAMUS Classification Assessment}\label{subsec:CAMUS_Classification}
We evaluate ED and ES classification with ResNet-18 and VGG-16. We compare real training data with real data augmented by unconditional, abstract token, SDM, or abstract token and segmentation images. Section~\ref{subsec:Experimental_Settings} defines the data balance, training protocol, metrics, and statistical analysis.

\begin{table}[!htbp]
\centering
\caption{Downstream ED/ES classification results using ACC, PR, RC, and F1. \rev{Values are means across five seeds. The $p$ column compares ACC with Real for the same backbone using the paired two sided Student t test.} Bold indicates the best results.}
\label{tab:classification_comparison}
{\footnotesize
\setlength{\tabcolsep}{3pt}
\begin{tabular}{@{}>{\raggedright\arraybackslash}p{0.3\textwidth}||ccccc@{}}
    \toprule
    \multirow{2}{*}{\textbf{Training data}} & \multicolumn{5}{c}{\textbf{Metrics}} \\
    \cmidrule(lr){2-6}
     & \textbf{ACC (\%)} $\uparrow$ & \textbf{$p$} & \textbf{PR (\%)} $\uparrow$ & \textbf{RC (\%)} $\uparrow$ & \textbf{F1 (\%)} $\uparrow$ \\
    \midrule
    \multicolumn{6}{c}{\textbf{ResNet-18}} \\
    \midrule
    Real & \rev{83.8} & \rev{N/A} & \rev{84.0} & \rev{83.8} & \rev{83.7} \\
    Real+Unconditional & \rev{84.2} & \rev{0.55} & \rev{84.4} & \rev{84.2} & \rev{83.9} \\
    Real+\rev{Abstract token} & \textbf{\rev{87.2}} & \rev{$<0.01$} & \textbf{\rev{87.4}} & \textbf{\rev{87.2}} & \textbf{\rev{87.0}} \\
    Real+SDM & \rev{83.4} & \rev{0.53} & \rev{83.6} & \rev{83.4} & \rev{83.3} \\
    Real+Abstract token and segmentation & \rev{86.3} & \rev{$<0.01$} & \rev{86.6} & \rev{86.3} & \rev{86.2} \\
    \midrule
    \multicolumn{6}{c}{\textbf{VGG-16}} \\
    \midrule
    Real & \rev{74.5} & \rev{N/A} & \rev{74.8} & \rev{74.5} & \rev{74.6} \\
    Real+Unconditional & \rev{75.1} & \rev{0.28} & \rev{75.3} & \rev{75.1} & \rev{74.9} \\
    Real+\rev{Abstract token} & \textbf{\rev{78.7}} & \rev{$<0.01$} & \textbf{\rev{78.9}} & \textbf{\rev{78.7}} & \textbf{\rev{78.5}} \\
    Real+SDM & \rev{75.9} & \rev{0.03} & \rev{76.3} & \rev{75.9} & \rev{75.6} \\
    Real+Abstract token and segmentation & \rev{77.7} & \rev{$<0.01$} & \rev{78.1} & \rev{77.7} & \rev{77.5} \\
    \bottomrule
\end{tabular}
}
\end{table}

\rev{Abstract token conditioning gives the highest mean accuracy for ResNet-18 (87.2\%) and VGG-16 (78.7\%).}

\subsection{CardiacUDC Segmentation and External Generalization}\label{subsec:CardiacUDC_segmentation}
\rev{We evaluate segmentation on the 227 CardiacUDC test frames in two settings. In the within dataset setting, we train and select all models on CardiacUDC. For the 2D U-Net, we compare Real with Real+50\%, Real+100\%, and Real+200\%. We also compare Real and Real+100\% using nnU-Net~\cite{isensee2021nnu}. These models predict LV, LA, RA, and RV.}

\rev{For direct external evaluation, training and checkpoint selection use CAMUS 4CH images. We retain the two labels shared with CardiacUDC, the LV cavity and LA. For both architectures, we compare five paired runs using Real or Real+100\%, where Real+100\% adds one synthetic image for each real CAMUS training image. We evaluate each selected CAMUS checkpoint directly on CardiacUDC.}

\begin{table}[!htbp]
\centering
\caption{\rev{CardiacUDC segmentation results using Dice, HD, and ASD. Panel (a) trains and selects models on CardiacUDC. Panel (b) trains and selects models on CAMUS and tests them directly on CardiacUDC. Values are means across five runs. Each $p$ column compares mean Dice with Real for the same architecture. HD and ASD are in pixels at $256 \times 256$ resolution.}}
\label{tab:cardiacudc-segmentation}
{\footnotesize\rev{\textbf{(a) Within dataset CardiacUDC evaluation}}\par}
\par\smallskip
{\footnotesize
\setlength{\tabcolsep}{2pt}
\renewcommand{\arraystretch}{0.85}
\resizebox{\textwidth}{!}{%
\begin{tabular}{@{}lcccccc|ccccc|ccccc@{}}
\toprule
\multirow{2}{*}{\textbf{Training data}} & \multicolumn{6}{c|}{\textbf{Dice (\%)} $\uparrow$} & \multicolumn{5}{c|}{\textbf{HD} $\downarrow$} & \multicolumn{5}{c}{\textbf{ASD} $\downarrow$} \\
\cmidrule(lr){2-7} \cmidrule(lr){8-12} \cmidrule(lr){13-17}
 & \textbf{Mean} & \textbf{$p$} & \textbf{LV} & \textbf{LA} & \textbf{RA} & \textbf{RV} & \textbf{Mean} & \textbf{LV} & \textbf{LA} & \textbf{RA} & \textbf{RV} & \textbf{Mean} & \textbf{LV} & \textbf{LA} & \textbf{RA} & \textbf{RV} \\
\midrule
\multicolumn{17}{c}{\rev{\textbf{2D U-Net}}} \\
\midrule
Real & \rev{79.7} & \rev{N/A} & \rev{81.7} & \rev{75.6} & \rev{82.7} & \rev{78.6} & \rev{17.7} & \rev{16.9} & \rev{20.7} & \rev{18.6} & \rev{14.7} & \rev{3.7} & \rev{3.4} & \rev{3.9} & \rev{3.6} & \rev{4.0} \\
Real+50\% & \rev{80.2} & \rev{$<0.01$} & \rev{82.2} & \rev{76.4} & \rev{83.1} & \rev{79.2} & \rev{16.2} & \rev{15.9} & \rev{18.7} & \rev{16.8} & \rev{13.4} & \rev{3.4} & \rev{3.1} & \rev{3.6} & \rev{3.3} & \rev{3.6} \\
Real+100\% & \rev{80.7} & \rev{$<0.01$} & \rev{82.7} & \rev{77.1} & \rev{83.5} & \rev{79.6} & \rev{15.4} & \rev{15.2} & \rev{17.5} & \rev{15.9} & \rev{13.0} & \rev{3.1} & \rev{2.9} & \rev{3.3} & \rev{3.0} & \rev{3.2} \\
Real+200\% & \rev{80.8} & \rev{$<0.01$} & \rev{82.8} & \rev{77.0} & \rev{83.5} & \rev{79.7} & \rev{15.1} & \rev{14.9} & \rev{17.1} & \rev{15.6} & \rev{12.8} & \rev{3.0} & \rev{2.8} & \rev{3.2} & \rev{2.9} & \rev{3.1} \\
\midrule
\multicolumn{17}{c}{\rev{\textbf{nnU-Net}}} \\
\midrule
\rev{Real} & \rev{84.0} & \rev{N/A} & \rev{86.3} & \rev{80.0} & \rev{86.5} & \rev{83.2} & \rev{13.8} & \rev{13.1} & \rev{16.2} & \rev{14.0} & \rev{11.9} & \rev{2.7} & \rev{2.5} & \rev{2.9} & \rev{2.7} & \rev{2.7} \\
\rev{Real+100\%} & \rev{84.6} & \rev{0.01} & \rev{86.8} & \rev{80.9} & \rev{87.0} & \rev{83.8} & \rev{12.8} & \rev{12.2} & \rev{14.8} & \rev{13.1} & \rev{11.1} & \rev{2.4} & \rev{2.3} & \rev{2.6} & \rev{2.4} & \rev{2.3} \\
\bottomrule
\end{tabular}%
}
\par\smallskip
{\footnotesize\rev{\textbf{(b) CAMUS to CardiacUDC external evaluation}}\par}
\par\smallskip
\setlength{\tabcolsep}{3pt}
\begin{tabular}{@{}lcccccc@{}}
\toprule
\multirow{2}{*}{\textbf{Training data}} & \multicolumn{4}{c}{\textbf{Dice (\%)} $\uparrow$} & \multirow{2}{*}{\textbf{HD} $\downarrow$} & \multirow{2}{*}{\textbf{ASD} $\downarrow$} \\
\cmidrule(lr){2-5}
 & \textbf{Mean} & \textbf{$p$} & \textbf{LV} & \textbf{LA} & & \\
\midrule
\multicolumn{7}{c}{\rev{\textbf{2D U-Net}}} \\
\midrule
\rev{Real} & \rev{70.8} & \rev{N/A} & \rev{76.9} & \rev{64.7} & \rev{27.6} & \rev{5.9} \\
\rev{Real+100\%} & \rev{71.6} & \rev{0.02} & \rev{77.5} & \rev{65.7} & \rev{26.5} & \rev{5.6} \\
\midrule
\multicolumn{7}{c}{\rev{\textbf{nnU-Net}}} \\
\midrule
\rev{Real} & \rev{74.6} & \rev{N/A} & \rev{80.2} & \rev{69.0} & \rev{22.9} & \rev{4.6} \\
\rev{Real+100\%} & \rev{75.2} & \rev{0.03} & \rev{80.7} & \rev{69.7} & \rev{22.1} & \rev{4.4} \\
\bottomrule
\end{tabular}
}
\end{table}

\rev{Within CardiacUDC, augmentation increases mean Dice from 79.7\% to 80.8\% for 2D U-Net and from 84.0\% to 84.6\% for nnU-Net, with lower mean HD and ASD in both comparisons. Under direct CAMUS to CardiacUDC evaluation, Real+100\% increases mean Dice from 70.8\% to 71.6\% for 2D U-Net and from 74.6\% to 75.2\% for nnU-Net.}

\subsection{CAMUS Segmentation Failure Cases}\label{subsec:Failure_Case_Analysis_of_Downstream_Tasks}
\begin{figure}[!htbp]
    \centering
    \includegraphics[width=0.48\textwidth]{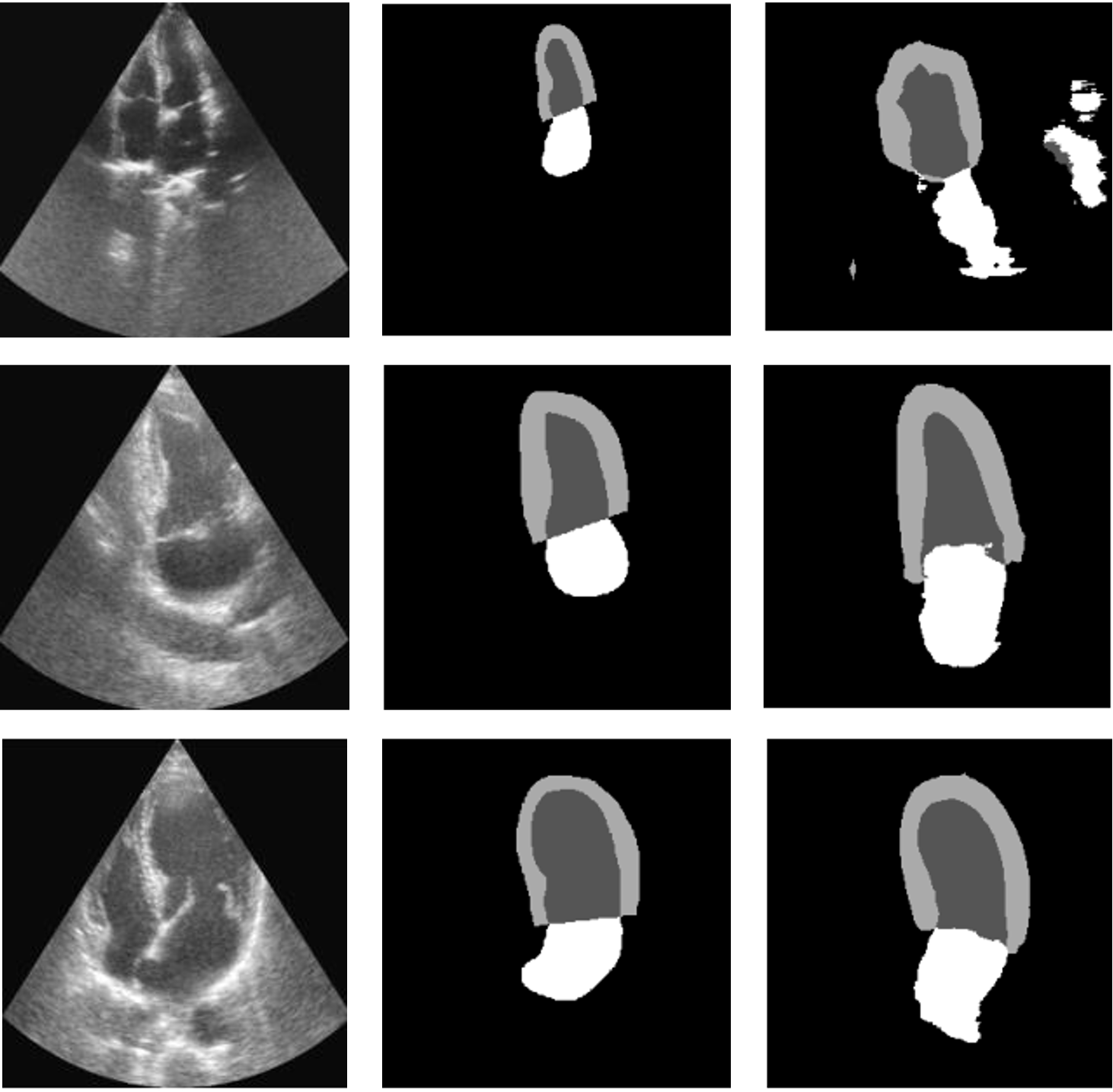}
    \caption{\rev{CAMUS segmentation errors for the model trained with Real+100\%. Columns from left to right show the image, reference mask, and predicted mask.}}
	\label{fig:segmentation_failure_cases}
\end{figure}

We inspect three segmentation errors from the CAMUS model trained with Real+100\%. Figure~\ref{fig:segmentation_failure_cases} shows the image, reference mask, and prediction for each case.
The first prediction contains several regions outside the reference LV endocardium and LA. The second and third predictions show local deviations at the LA and LV endocardium boundaries.

\section{Discussion}
\label{sec:Discussion}
The model conditioned on an abstract token and segmentation mask has the lowest mean FID of the five generation settings in Section~\ref{subsec:Image_Synthesis_Evaluation} and Table~\ref{tab:inception_scores}.

\rev{The controlled sampling experiment in Section~\ref{subsec:Text_Condition_Contribution} and Table~\ref{tab:text-contribution} isolates the role of the abstract token while retaining the segmentation mask and trained model. The abstract token and mask provide complementary information. The token specifies view dependent anatomy that is absent from the CAMUS maps, while the map preserves the annotated geometry.}

\rev{The prompt experiment in Section~\ref{subsec:Prompt_Ablation} shows similar synthesis and classification results for plain English and abstract tokens in Table~\ref{tab:prompt-ablation}. In the classification experiment in Section~\ref{subsec:CAMUS_Classification}, abstract token augmentation performs best for both backbones in Table~\ref{tab:classification_comparison}. Together, these results indicate that class identity supplies most of the useful prompt information for the evaluated task. Natural language wording does not improve classification over abstract tokens in this experiment.}

The CAMUS segmentation experiment in Section~\ref{subsec:Downstream_tasks} shows modest gains from synthetic augmentation in Table~\ref{tab:dice-hd-asd}. \rev{The largest synthetic set gives no additional mean Dice gain after the synthetic set matches the size of the real training set. The regional HD pattern indicates that the benefit is concentrated in the LA rather than uniform across structures.}

\rev{The CardiacUDC experiments in Section~\ref{subsec:CardiacUDC_segmentation} and Table~\ref{tab:cardiacudc-segmentation} show modest augmentation gains after training within a second dataset, with nnU-Net, and under direct evaluation from CAMUS to CardiacUDC. Similar gains occur on the CardiacUDC test split, under direct external evaluation, and with both segmentation architectures.}

\rev{Table~\ref{tab:dice-hd-asd} reports CAMUS validation performance from the same split used for checkpoint selection. The CAMUS results therefore compare training conditions within this split. Table~\ref{tab:cardiacudc-segmentation} reports test performance after selection on separate CardiacUDC validation frames and direct external test performance after training and selection on CAMUS. The CardiacUDC results assess the augmentation effect on unseen data.}

The errors in Section~\ref{subsec:Failure_Case_Analysis_of_Downstream_Tasks} occur near structure boundaries and in small regions in Fig.~\ref{fig:segmentation_failure_cases}. \rev{These cases suggest that future augmentation should target variation near structure boundaries and in small regions.}

\section{Conclusion}
\label{sec:Conclusion}
In this paper, we studied three diffusion architectures for echocardiography image synthesis. These include unconditional generation, text conditioned generation, and generation with an abstract token and segmentation mask. \rev{The model conditioned on an abstract token and segmentation mask has the lowest mean FID of the five evaluated settings. The controlled sampling experiment shows that the abstract token specifies right heart anatomy that is absent from the CAMUS maps. Synthetic augmentation gives modest segmentation gains within CAMUS and CardiacUDC, including direct evaluation from CAMUS to CardiacUDC and evaluation with nnU-Net. On CAMUS, mean Dice does not increase from Real+100\% to Real+200\%, and the HD reductions are largest for the left atrium.}

\subsection*{Data and Materials Availability}
The data that support the findings of this study are available from public sources. The CAMUS echocardiography dataset is available from the \href{https://www.creatis.insa-lyon.fr/Challenge/camus/}{CAMUS project website} and is described by Leclerc et al.~\cite{leclerc2019deep}. \rev{CardiacUDC is available from the} \href{https://www.kaggle.com/datasets/xiaoweixumedicalai/cardiacudc-dataset}{\rev{CardiacUDC release}} \rev{and is described by Yang et al.~\cite{yang2023graphecho}. The study uses public deidentified datasets and involves no new patient data collection.}

\subsection*{Acknowledgments}
We acknowledge the support of the Natural Sciences and Engineering Research Council of Canada (NSERC), funding reference number RGPIN-2023-03575.

\bibliographystyle{spiejour}
\bibliography{PaperContent/References/ref}

\end{document}